# Broadening the scope of weak quantum measurements *II*: Past and future measurement effects within a double Mach-Zehnder-interferometer setting

Yakir Aharonov[1], Eliahu Cohen[1], Avshalom C. Elitzur[2]

*Following earlier applications of weak measurement to new cases (Part I), we proceed to explore its temporal peculiarities. We analyze an idealized experiment in which weak which-path measurements do not prevent consecutive weak interference effects, and then again the which-path information is recovered by strong measurements. We also show how the same effect can be obtained even by all these measurements being carried out on a single particle. The simplified setting enables critically assessing competing interpretations of the results. The most natural one is that of the Two-State-Vector Formalism, according to which the quantum "two-state" between an earlier and a later measurement is equally determined by both of them.*

**Key Words**: weak measurement, TSVF, which-path, weak values.

## Introduction

Weak measurement [1-7], which has proved effective in revealing unique quantum values hitherto believed to be unobservable or even nonexistent [2-5], has been devised within the Two-State-Vector Formalism (TSVF), based in turn on the assumption that quantum interaction involves a combination of past and future state vectors. In this Part *II* of our article we prove this unique assertion with the aid of a simple but delicate experiment.

[1] *School of Physics and Astronomy, Tel Aviv University, Tel Aviv 69978, Israel*

[2] *Iyar, The Israeli Institute for Advanced Research, Rehovot, Israel*

Recently an analogous experiment [8], utilizing weak quantum measurements has demonstrated consecutive "which-path" and interference measurements within one and the same double-slit experiment . A comprehensive discussion of this setting has been made by [9] from the viewpoint of the TSVF. In what follows we analyze these results within an idealized setup, along lines similar of a previous work [7] and in keeping with some TSVF's critics [*e.g.,* 10], who urge phrasing it in a more lucid manner. We show how which-path results can be obtained while leaving interference intact, and then again be fully "resurrected" after interference. Next we present a single-photon version (see Part *I*) of this experiment. Finally we consider an alternative account and show TSVF to be superior.

## 1. TSVF AND WEAK VALUES

The foundations of time-symmetric QM were laid by Aharonov, Bergman and Lebowitz [1]. The probability for measuring the eigenvalue $c_j$ of an observable C, given the initial and final states $|\psi(t')\rangle$ and $\langle\Phi(t'')|$, respectively, can be described by the time-symmetric formula:

$$(1) \quad P(c_j) = \frac{|\langle\Phi(t'')|c_j\rangle\langle c_j|\psi(t')\rangle|^2}{\sum_i |\langle\Phi(t'')|c_i\rangle\langle c_i|\psi(t')\rangle|^2}.$$

It was later realized [2-3] that this probability is not a mere counterfactual. Rather, it can be measured, offering novel results via the specific operator O's weak value:

$$(2) \quad O_W = \frac{\langle\Phi|O|\Psi\rangle}{\langle\Phi|\Psi\rangle}.$$

Performing a weak measurement in the sense of [7] enables avoiding the uncertainty disturbance inflicted on the outcome and achieving the weak value using the pointer's deviation: By probability theory's Large Numbers Law, the above slight deviation is highly significant when summed over a sufficiently large *N*. Hence, *should the weak measurement be followed by a strong one, the latter would always confirm the former's result.*

This is the background for the feat we want to present, namely, quantum measurement of simultaneous noncommuting values.

## 2. WHICH-PATH AND INTERFERENCE

Consider then a photon passing through a system of two consecutive Mach-Zehnder Interferometers (Fig. 1). Suppose first that it undergoes no which-path measurement. It will traverse $MZI_1$ superposed on both $L_1$ and $R_1$. Next it will go through $MZI_2$ only on $R_2$ (which is $MZI_1$'s interference exit), finally exiting from $MZI_2$ with equal probability towards $L_f$ and $R_f$.

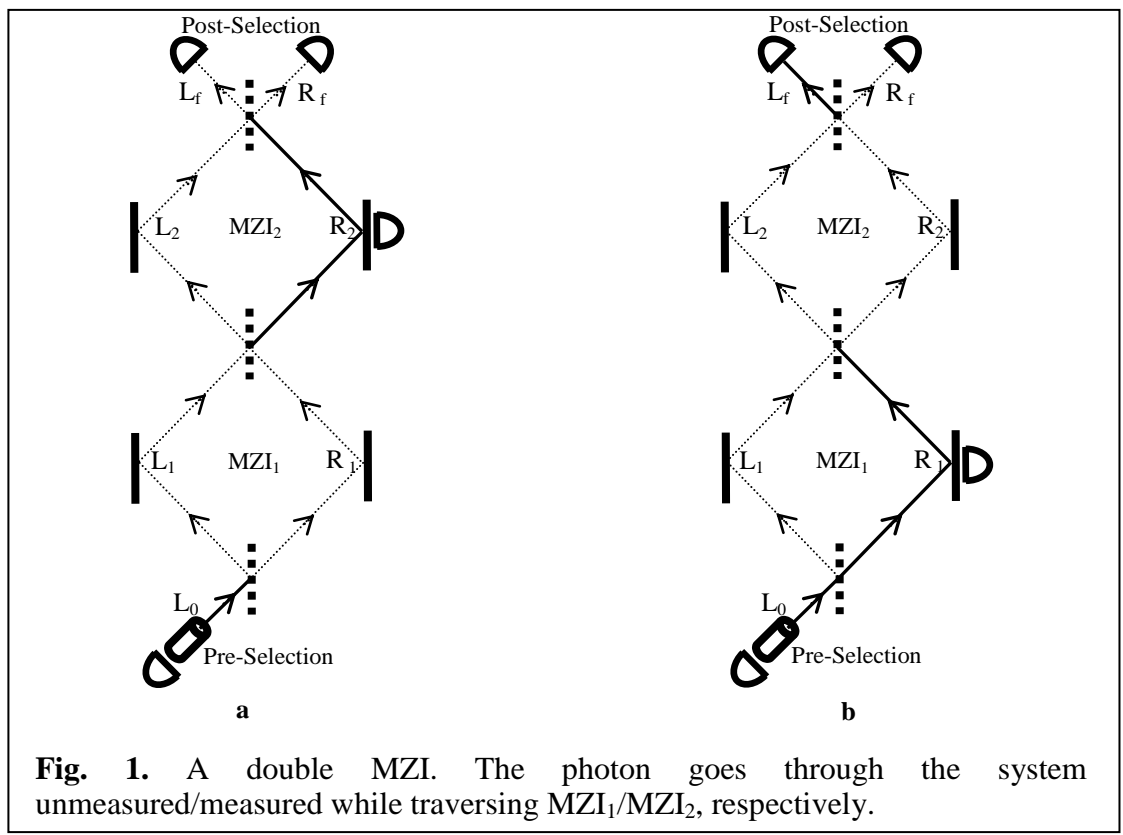

**Fig. 1.** A double MZI. The photon goes through the system unmeasured/measured while traversing $MZI_1$/$MZI_2$, respectively.

If, however, the photon is measured along $MZI_1$, it will take either $L_1$ or $R_1$. Consequently, it will traverse $MZI_2$ superposed, *on both sides*. Finally, while interference has been disturbed in $MZI_1$, it re-emerges in the two final measurements, whose $L_f$/$R_f$ outcomes match the earlier $R_1$/$L_1$ ones. In other words, refraining from which-path measurement on MZI$_2$ enables resurrecting the which-path information from MZI$_1$.

These familiar tradeoffs, derived from the uncertainty principle,

(3) $\Delta x \Delta p \geq \dfrac{\hbar}{2}$,

take a new twist when the measurements are weak.

### 3. WEAKENING THE MEASUREMENT

The essential elements of both ordinary and weak measurement have been described in an earlier work [7] with the aid of a Michelson interferometer. In ordinary measurement, one of the reflecting mirrors can be turned into a detector by appropriately increasing its position uncertainty so as to increase its momentum precision. To make the measurement weak, the mirror should be allowed lower position uncertainty, such that its coupling to the photon's momentum is weakened.

This measurement is unique in that, when a large number of its outcomes is summed up, their collective precision goes up to that of ordinary measurement. This summation, however, works differently for the disturbance caused by weak measurement. The relative error, over a large ensemble, approaches 0.

In [7] we have also advocated two technical features that enhance weak measurement. First, for which-path measurements, using a single detector rather than two reduces the chance for collapse. Second, rather than letting the detector interact with all $N$ particles and then indicate their collective impact, it is better to separately record each individual outcome, as imprecise as it must be, calibrating the detector anew before the next. This method enables "slicing" the individual outcomes, e.g., re-grouping them for more accurate summation.

## 4. COMBINING STRONG AND WEAK MEASUREMENTS

With the above in mind, let us combine strong and weak measurements within our double-interferometer setup (Fig. 2).

Consider an ensemble of photons going, one by one, through the setting, undergoing weak and strong measurements, henceforth marked with <u>underline</u> or **boldface**, respectively. A **strong** measurement is already performed by the photon's very emission from the bottom-left corner. Second comes a <u>weak</u> which-path measurement of the photon's path through MZI$_1$ (symbolized by the gray detector), followed by another <u>weak</u> measurement through MZI$_2$. Finally, a **strong** measurement is performed by the last two detectors on L$_f$ and R$_f$.

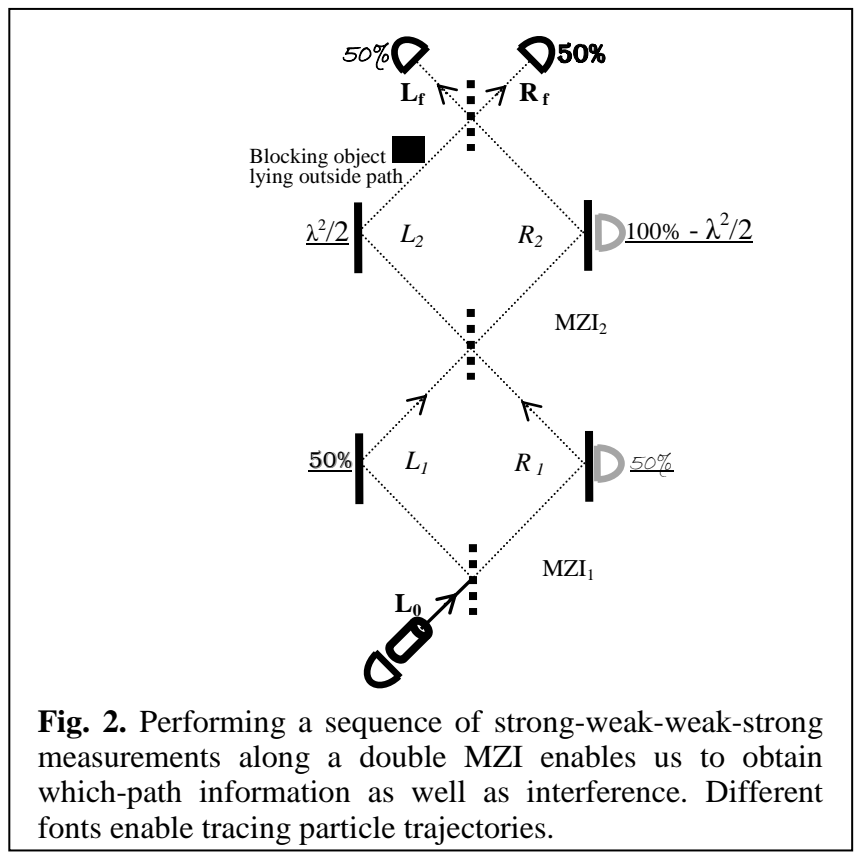

**Fig. 2.** Performing a sequence of strong-weak-weak-strong measurements along a double MZI enables us to obtain which-path information as well as interference. Different fonts enable tracing particle trajectories.

Have all measurements been **strong**, the price exerted by the uncertainty principle would be as described in the previous section. Now, however, it is the weak "which-path" measurement that gives <u>50%L$_1$</u>-<u>50%R$_1$</u>. This may be interpreted as indicating that each particle has taken *either* path, but the case is much more peculiar: The next weak measurement, in MZI$_2$, gives ~0%L$_2$:~100%R$_2$, preserving the initial strong **"right-up"** momentum, just as it would do with the MZI$_2$ measurement being strong! This interference pattern indicates that each photon has traversed *both* L$_1$ and R$_1$, *superposed*. In other words, we seem to have measured the *wave-function* without "collapsing" it [7].

The next surprise is brought by the final measurements. Having obtained and recorded each weak outcome separately, we now have the freedom to slice these outcomes into any sub-ensembles as we choose. This is, in fact, a pre- and post-selection process, giving rise, respectively, to the initial ("preparation") and final states.

So, upon obtaining the *N* final **L$_f$** and **R$_f$** outcomes (marked in Fig.2 with distinct fonts to enable following their trajectories), we divide the earlier MZI$_1$ outcomes accordingly. Summing again each sub-ensemble separately, the entire *N*'s 50%L$_1$:50%R$_1$ now gives its place to a slight but significant bias #(L$_1$)>#(R$_1$) and #(L$_1$)<#(R$_1$) for each *N/2*. Lo and behold, the outcomes of the initial and final measurements match at probability $P \to 1$ as the ensemble grows.

Indeed, the weak values of the projection operators $\pi_{L1} \equiv |L_1\rangle\langle L_1|$ and $\pi_{R1} \equiv |R_1\rangle\langle R_1|$, can be found according to [5] :

$$(4) \quad \langle \pi_{L1} \rangle_w = \frac{\langle R_f | \pi_{L1} | L_0 \rangle}{\langle R_f | L_0 \rangle} = \frac{0.5}{0.5} = 1 \quad , \quad \langle \pi_{R1} \rangle_w = \frac{\langle R_f | \pi_{R1} | L_0 \rangle}{\langle R_f | L_0 \rangle} = \frac{0}{0.5} = 0 \quad ,$$

in the case of a perfect weak measurement. The affinity between conditional probability and time-symmetric interpretations was explained in [1] and indeed, these results can be achieved using basic probabilistic rules.

This result's oddity is obvious. The interference observed in MZI$_2$ is supposed to amount to quantum erasure, namely, "forgetting" of MZI$_1$'s which-path information. And yet, the final measurement resurrects the latter.

## 5. A SINGLE-PHOTON VERSION

Weak measurement requires a sufficiently large *N* outcomes to be summed up. Yet we have shown [7] that this ensemble does not have to be of *N* particles, but rather of *N* weak measurement outcomes of *one and the same particle*. Such a single-photon variant of our experiment is shown next.

Let our single photon bounce back and forth through our double-MZI system (Fig. 4). Let the solid mirrors at the beginning and end act also as ordinary detectors in the sense of Sec.2. We note that this setting requires high accuracy and minimal energy losses, otherwise the photon would collapse before the experiment's end.

Being strongly measured at the beginning and end of each cycle, the photon's starting points vary with equal probability between $L_0$ and $R_0$, and between $L_f$ and $R_f$.

Here again, in all cases where the photon has begun from $L_0/R_0$ and ended at $L_f/R_f$ or *vice versa*, the weak measurements on $MZI_1$ and $MZI_2$ equally agree with the past and future strong measurements: **$R_1$**-$\underline{L_f}$, **$L_1$**-$\underline{R_f}$, **$R_2$**-$\underline{L_o}$, **$L_2$**-$\underline{R_o}$.

And here again, these pairs of measurements, manifesting interference due to the momentum left undisturbed, do not disturb the *position* measurements ($\underline{L/R_2}$ and $\underline{L/R_1}$, respectively) which took place between them.

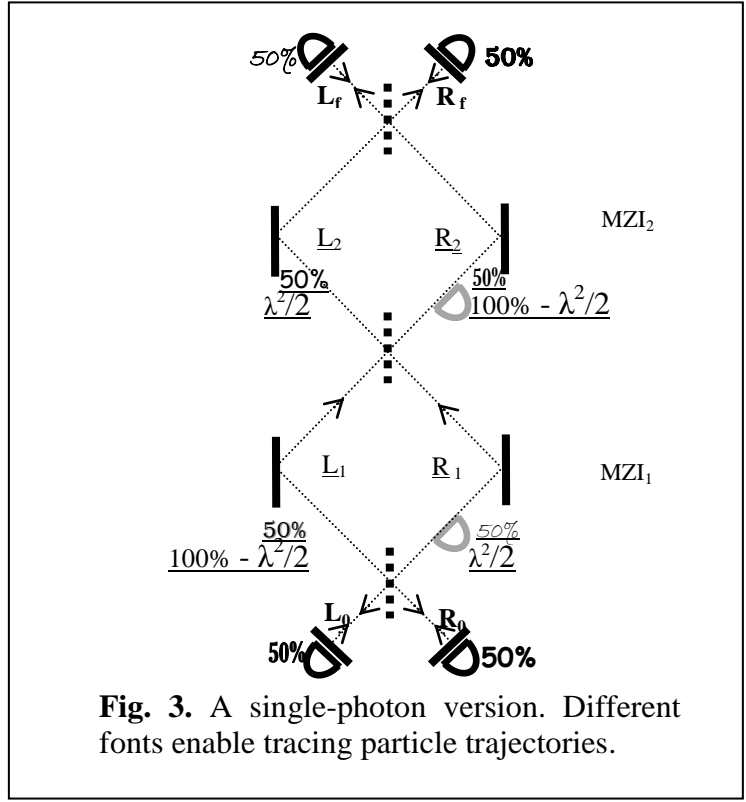

**Fig. 3.** A single-photon version. Different fonts enable tracing particle trajectories.

## 6. CONSIDERING ALTERNATIVE HYPOTHESES

Naturally, if Sec. 4's results can be accounted for by a more orthodox account, it merits consideration.

Such a possibility seems to be offered by the interference shown to be *imperfect*. Have the $MZI_1$ which-path measurement been strong, $L_2$ would be traversed by **0** photons. But as it is weak, then, by שגיאה! מקור ההפניה לא נמצא, approximately $\lambda^2/2$ photons traverse $L_2$. These are the few photons that have been collapsed by the weak measurement to this side. Moreover, Tollaksen *et al.* [9] showed that, in order for the weak measurements to work, the $L_2$ path must remain open. Blocking it (black square in Fig. 2), would ruin the above resurrection. Is it possible, then, that these few collapsed photons do the entire trick?

Our slicing method derives from this account the following prediction: Slicing the $\underline{L}_2$ outcomes according to accurate (AC) and inaccurate (IN) outcomes, i.e., $\underline{L}_1$-**R**$_f$ and $\underline{R}_1$-**L**$_f$ matches and mismatches, should give

$$(5) \quad P(R_1 | L_f, AC) = \frac{\lambda^2}{2N}$$

$$(6) \quad P(R_1 | L_f, IN) = \frac{N - \lambda^2}{2N},$$

which can be straightforwardly predicted to fail. Suppose we use only 1/10 of the outcomes. Since λ can be small, *e.g.*, 2 or 3, it is quite likely that the *N/10* particles do not include these $\lambda^2$ collapsed photons. However, *N/10* being still very large, we expect the same resurrection of the L$_1$/R$_1$ outcomes within these sub-ensembles too.

In fact, the "$\lambda^2$ alternative" is ruled out even more straightforwardly by the single photon version (Sec. 5 above), where a single collapse can ruin the entire experiment.

The above alternative is a specific example of a "one-vector account." This category includes almost all prevailing interpretations of QM. It is possible, *e.g.*, to argue that the weak value "was there all along," the measurement merely revealing it [11]. Other alternatives are offered by "no collapse" [12] or super-deterministic interpretations.

Competing interpretations of QM, by their giving the same predictions, are widely agreed to be a matter of philosophical inclination. It should be borne in mind, however, that the very idea of weak measurement has been devised especially for testing the unique predictions of the TSVF (among which the present paper gives only one example). TSVF, therefore, is the most natural framework for understanding its own unique predictions, especially the "super-weak" ones [13].

## 7. DISCUSSION

The TSVF presents two sets of results whose combination is considered impossible within orthodox quantum theory, *i.e.*, "having our cake and eating it too" by observing interference while resurrecting the earlier which-path information. This feat can be applied to any other pair of noncommuting operators. It can also be demonstrated on a single particle.

The slicing method enables critically assessing the weak measurement's results and considering rival explanations. As the soundest explanation we have opted for TSVF's complementary feat in the realm of theorizing: A state between two consecutive measurements can be best understood as carrying *both* their effects, namely, being superposed and, *at the same time*, assuming a definite value. That the resulting state exhibits even more surprising features, such as those presented in consecutive works [14-15], is only natural.

**Acknowledgements**

We thank Paz Beniamini, Shay Ben-Moshe, Einav Friedman, Robert Griffiths and Doron Grossman for helpful comments and discussions.

This work has been supported in part by the Israel Science Foundation Grant No. 1125/10.